**Special Relativity Teaches the One-Way Velocity of Light Is Measurable Without Synchronized Clocks and Clocks Are Synchronizable Without One-Way Signals**

Charles Nissim-Sabat[*]

Northeastern Illinois University (*emer.*)

Chicago, IL 60625 U.S.

Recent authors have addressed the Reichenbach assertions that the one-way velocity of light, OWVL, cannot be measured because we lack a method to synchronize distant clocks, and that one needs OWVL to synchronize distant clocks. Thus, one may have any OWVL provided it is greater or equal to $c/2$ and the round-trip velocity is $c$, and, since clocks can be synchronized only by light signals, clock synchronization depends on whatever OWVL happens to be. I show that the Dragowski and Wlodarczyk proposal to use transverse Doppler shifts to measure one-way velocities cannot succeed. In fact, OWVL is measurable in a synchronization-free manner by determining the minimum wavelength transmitted by an airgap optical barrier. OWVL along three non-coplanar axes is required to determine that OWVL is $c$. Also, contrary to the Thyssen claim that distant simultaneity is conventional, I show that, using a moving clock, relativistic time-dilation allows the synchronization of isochronized stationary clocks without knowing OWVL and also allows, separately, the measurement of OWVL. A burst of unstable particles is a suitable moving clock. Thus, one-way velocity and synchronization are shown to be independent concepts. The wavelength transmitted by an airgap and the decay of unstable particles have been neglected regarding OWVL, but all experiments in these areas have found the one-way velocity of light is $c$.

Keywords: one-way velocity of light; clock synchronization; simultaneity; conventionality thesis.

[*]nissimsabatc@gmail.com

# 1. Introduction.

Recently Drągowski and Włodarczyk[1] have proposed to use the transverse Doppler effect to measure the one-way velocity of light, $c(\theta)$, which, if possible, would allow the synchronization of distant clocks and thus negate Thyssen's claim[2] that distant simultaneity is conventional.

Numerous experiments confirm predictions of Special Relativity ('SR'). The 'Postulate' of SR is often stated as:

*For all inertial observers,* (a) *the laws of physics are the same, and*

(b) *the velocity of light, c, is the same.*

There is controversy about (b): Is it "the velocity of light" for a **one-way** trip between two points or that for a **round-trip**?

When Einstein[3] introduced SR and derived the Lorentz transformations, he stipulated that, given two clocks A and B at rest with respect to each-other in field-free space, if a light *signal* is emitted at A at time $T_o$, reflected at B back to A, and received there at time $T_f$, the clocks are synchronized by setting the time when the signal reached B as $T_B = T_o + (1/2)(T_f - T_o)$. Einstein's synchronization thus assumed that the one-way velocity of light, $c(\theta)$, from A to B equals that from B to A, $c(\theta+\pi)$, both one-way velocities equaling $c$.

Reichenbach[4] argued that Einstein had no experimental or theoretical support for his assumption. All we know is that

$$T_B = T_o + \varepsilon(T_f - T_o) \tag{1a}$$

with $0 \leq \varepsilon \leq 1$, and one may **have or choose** $c(\theta)$ to be $c_+$ in an arbitrary direction and $c_-$ in the opposite direction, if, over any distance $L$,

$$(L/c_+) + (L/c_-) = (2L/c) \tag{1b}$$

And, therefore, Einstein's procedure is *an arbitrary convention.* Thus, Reichenbach enunciated a

*Conventionality of Synchronization ('CS') Thesis* (some authors speak of *Conventionality of Simultaneity)*:

(A) measuring the one-way velocity of light requires synchronizing two clocks a distance apart;

(B) clock-synchronization cannot be achieved without using the one-way velocity of light;

(C) whatever the one-way velocity of light happens to be, or chosen to be, according to Equation (1a), all the predictions of SR are satisfied.

Thus, tacitly, Reichenbach modified, clause (b) of Einstein's SR Postulate to read:

"*(b') the **two-way** velocity of light is c along any direction.*"

The CS thesis has been elaborated by Edwards;[5] Winnie,[6, 7] who derived synchronization-dependent Lorentz transformations consistent with those of SR and argued that every one-way velocity is conventional; Giannoni,[8] who showed the laws of classical physics are independent of $\varepsilon$; and many other philosophers and physicists, notably Grünbaum,[9] Ungar,[10] Tooley,[11] and Jammer.[12]

Yet, now, some philosophers argue it is nonsensical to speak of speed. "To understand Relativity, we have to expunge all ideas of things having speeds, including light," Maudlin.[13] But, in the next paragraph, Maudlin discusses "two [things] in very rapid relative motion." Einstein and Reichenbach would agree that, in vacuo, nothing is more rapid than light. No experimental approach that can measure $c(\theta)$ or synchronize clocks has been formulated, Anderson et al.[14]

I designate any electromagnetic or gravitational radiation as 'light.' Both radiations constitute transverse waves that transmit, in vacuo, energy, linear momentum, angular momentum, and information. I consider waves of frequency $v$ and round-trip wavelength in vacuo $\lambda = c/v$.

I will show that one can measure $c(\theta)$ without synchronizing clocks and synchronize clocks without measuring $c(\theta)$, that all experiments in two areas that have been neglected regarding $c(\theta)$ have all found $c(\theta) = c$, and that, most probably, systematic research will determine that $c(\theta) = c$.

In Section 2, I show that the determination that $c(\theta)$ for light-signals is always $c$ requires systematic synchronization-independent bi-directional measurements along three non-coplanar axes and why experimenters, Drągowski and Włodarczyk[1] among them, who have claimed to have found ways to measure $c(\theta)$ in fact have failed to do so.

In Section 3, I show that, with only one clock, $c(\theta)$ can be measured by experiments in frustrated total internal reflection by an airgap optical barrier, an area that has been ignored in this regard.

In Section 4, I disprove Thyssen's claim[2] that distant simultaneity is conventional by showing that SR's time-dilation teaches a conventionality-free method for synchronizing the two or more stationary clocks by using an isochronous moving clock. A burst of unstable particles extracted from an accelerator constitutes a suitable moving clock. Separately, one can determine the one-way velocities of the moving clock or of the particles. *Isochrony* of the stationary clocks yields their *synchronization*. Thus, clock synchronization and the one-way velocity of light are independent concepts.

## 2. On Drągowski and Włodarczyk and the Conventionality of Synchronization Thesis.

This Section uses only one clock: synchronization is irrelevant. To describe the *reciprocal asymmetry* between $c_+$ and $c_-$ I choose a *synchronization parameter* $\sigma$, $\sigma = 1 - 2\varepsilon$, with $0 \leq \sigma \leq 1$, $c_+$ being the maximum $c(\theta)$, and Eq. 1b being satisfied:

$$c_+ = c/(1-\sigma), \tag{2a}$$

$$c_- = c/(1+\sigma). \tag{2b}$$

*2.1 The CS Thesis Provides a Sole Expression for the Anisotropy in the One-Way Velocity of Light.*

Reichenbach[4] formulated the CS thesis as he sought to present a philosophical view of space and time in the light of SR. He emphasized what is measured, distinguishing this from what is assumed, such assumptions being often made for the sake of simplicity. Before SR, all space was thought to be filled by the ether and the velocity of light was determined by the properties of the ether just as the velocity of sound on earth is determined by the properties of air. SR eliminated the ether, thus leaving light propagating through a vacuum with no properties relevant to the propagation of light. Thus, Reichenbach conceived of geometry as a relation between points and of light as a means to study this relation, his conclusion being that the empirical evidence is that this relation is described by Euclidean geometry. *Arguendo,* I shall adopt his conclusions.

Given points A and B, the only conceivable experiment an observer at A can do is to measure the A to B distance and the time A's clock records between when A sends a signal towards a mirror at B and when the signal returns after being reflected by that mirror. Reichenbach assumed the speed of light is uniform from A to B and, separately, from B to A. From the A to B round-trip time and distance, the observer could choose $\varepsilon$ for Equation (1a) and formulate the CS thesis as stated in Section 1, with all its predictions duplicating those of Einstein's SR.

Figure 1 shows, in vacuo, in field-free space on the X, Y plane, a right triangle formed by points A at (0, 0), B at (0, x), and C at *(x, y)*, $r=(x^2+y^2)^{1/2}$ being the distance AC, and *sec $\theta$ = r/x*. An observer at A has a light-source and a clock while B and C are mirrors. Eq. (2a) allows $c(\theta) = \infty$ from A to B and from A to C, but in this case no choice of *c($\theta$)* for C to B would satisfy the SR requirement that the time for an ACBA round trip equal the ABCA perimeter divided by *c*.

The requirement that CS duplicate all SR predictions can be addressed as follows. Let $c(\theta)=c/(1-\sigma)$, along the +X-axis, Equation (2a), along the Y-axis $c(\theta)=c,$ satisfying Eq. (1b) and, along CA, $c(\theta)= c/(1+ \sigma f(\theta))$. The ABCA round-trip time equals the perimeter divided by *c*:

$$(1-\sigma) x/c + y/c + (x \sec\theta)(1+\sigma f(\theta))/c = x/c + y/c + (x \sec\theta)/c \qquad (3)$$

yielding $f(\theta)=cos\theta$ and

$$c(\theta) = c/(1- \sigma cos\theta). \tag{4}$$

The same can be done with a congruent AB'C' triangle with B' at $(-x, 0)$ and C' at $(-x, -y)$ and one obtains Eq. (4) with $\theta$ measured with respect to the +X axis. Eq. (4) is consistent with the choice $c(\theta)=c$ along the Y-axis and can be obtained in the same manner for an arbitrary triangle. Note that Eq. (3) is obtained by using Reichenbach's assumptions with A, B, and C being at rest.

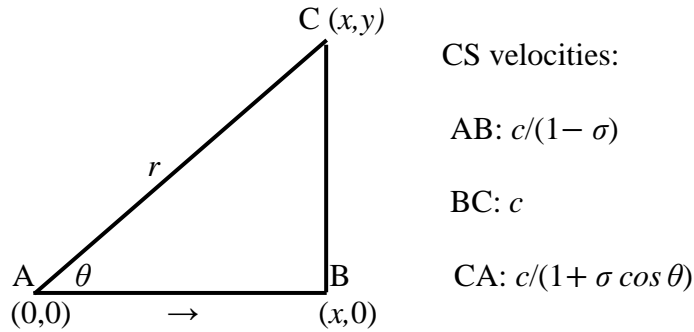

**Figure 1**. <u>Computation of the round-trip time for a light signal</u>. A clock at A measures the travel time for a signal from A to B to C and back to A.

If the signal from A has a round-trip velocity $v$, SR requires that the difference in round-trip times between this signal and the light signal be independent of $\sigma$ so that

$$v(\theta)= v/(1-\beta\sigma cos\theta) \tag{5}$$

or $v(\theta)= v/(1-\boldsymbol{\beta} \cdot \boldsymbol{\sigma})$, where $\boldsymbol{\beta} = \boldsymbol{v}/c$ and $\boldsymbol{\sigma}$ designates the maximum magnitude and direction of the asymmetry. If Equation (5) obtains, time (or phase) of arrival differences are independent of $c(\theta)$ for simultaneously-emitted light signals taking different paths or means between two points.

Consider again the ABC triangle with $c_r = c/(1+k)$ being a possible $c(\theta)$ along CA, $c(\theta)=c$ along BC, and $c/(1-\zeta)$ along AB. The round-trip time is:

$$T_{ABCA} = (r+y+x)/c + \{kx/\cos\theta) - \sigma x\}/c, \quad (6)$$

which, according to Equation (4), equals *(r+y+x)/c*. So *k= σcos θ*, *c(θ) = c/(1− σ cosθ)*. Equation (4) is the only expression for *c(θ)* that the CS thesis allows.

Finally, consider a triangle ADB congruent to the triangle ACB depicted in Figure 1 but extending in a plane XY' intersecting the XY plane at an arbitrary angle. The A source broadcasts simultaneously towards C and D and the CS thesis requires simultaneous return to A for these signals. Therefore *c(θ) = c/(1− σ cosθ)* applies to any arbitrarily chosen XY plane. Equation (3) azimuthally symmetric function fully describes what the CS thesis allows as the only possible anisotropy for *c(θ)* in three dimensions.

From the above, *c(θ) = c* along any axis Y' orthogonal to the X axis, i.e. everywhere along the whole YZ plane. Therefore, *c(θ)* is *c* in any direction orthogonal to the one where its anisotropy is, or is chosen to be, maximal and <u>one needs measurements along three non-coplanar axes to prove *c(θ)* is isotropic</u> and, for each axis, these measurements must be simultaneous in both directions so as to also test the 'R-limited' second clause of SR's Postulate.

The above conclusions are <u>independent of *σ* and of the line chosen for the X axis.</u>

*2.2 Difficulties in Designing Measurements of the One-Way Velocity of Light.*

<u>2.2.1 Kinematics.</u>

Relevance of Equations (4) and (5) to the analysis of several kinematical experiments is treated in detail by Nissim-Sabat[15] who has shown that one cannot determine *c(θ)* by measuring time or phase differences when two light signals are emitted simultaneously and then recombined after each signal passes through different media (glass, water, etc...) or generates a pulse in a cable.

This is often overlooked. For instance, Dryzek and Singleton[16] claim to have measured *c(θ)=c* for photons produced in opposite directions by positron annihilation at rest midway between two counters. The pulses from each counter are transmitted by cable to a coincidence circuit and the authors find maximum coincidences when the cables have equal length. They note the controversy concerning $c(\theta)$ for photon travel in air but they tacitly assume that $v(\theta)$ for signals along their cables is independent of direction. What they have shown is that Equations (4) and (5) are valid.

Thus, if $c(\theta) = c/(1- \sigma \cos\theta)$, $c(\theta)$ cannot be measured by arrival-time differences for light-signals but SR is valid. On the other hand, if $c(\theta) \neq c/(1- \sigma \cos\theta)$ in such an experiment, SR is not valid but one can measure $c(\theta)$ by systematic measurements of arrival-time differences for light-signal travel between two points along different paths. Yet, given the ubiquitous use of light-signals for land surveys, if $c(\theta) \neq c/(1- \sigma \cos\theta)$, it would have been observed.

2.2.2 Drągowski and Włodarczyk's suggested transverse Doppler-shift experiment.

Equation (5) explains why Roemer's method and Doppler shifts experiments fail to measure $c(\theta)$. Let, in Figure 1, segments AB, BC, and CB have lengths $r_A$, $r_B$, and $r_C$ and angles $\theta_A$, $\theta_B$, and $\theta_C$ with respect to the direction where $c(\theta)$ is maximal. A is a clock-counter combination and a source S with velocity $v(\theta_B)$ at B sends a signal to A, signal velocity $c(\theta_A)$, travel time $r_A/c(\theta_A)$. S proceeds to C, travel time $r_B/v(\theta_B)$, whence it sends a signal to A, travel time $r_C/c(\theta_C)$. From Equations (4) and (5), the difference in travel times $T_{CA} - T_{BA}$,

$$r_B/v(\theta_B) + r_C/c(\theta_C) - r_A/c(\theta_A) = r_B/v + r_C/c - r_A/c, \qquad (7)$$

is independent of the angles and of $v(\theta_B)$: Doppler shifts cannot measure $c(\theta)$. The same reasoning applies to many of the experiments that Salmon[17] has shown cannot measure $c(\theta)$.

In an arrangement they believe is different from that in Figure 1, Drągowski and Włodarczyk[1] have proposed that one use the transverse Doppler shift of the radiation emitted by a radioactive source to measure its one-way velocity $v(\theta_B)$ with respect to the observer. Figure 2 illustrates their proposal.

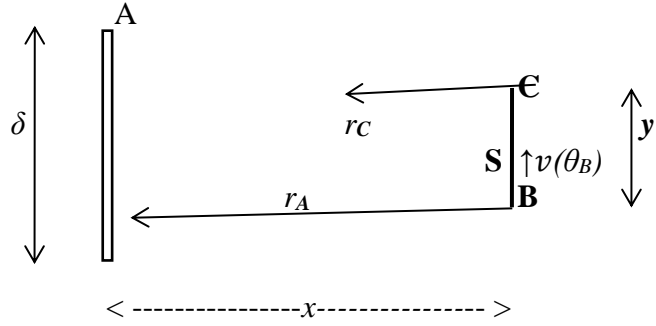

**Figure 2**. <u>Transverse Doppler Shift</u>. **S**, a source with known round-trip velocity $v$ and unknown one-way velocity $v(\theta_B)$, emits rays $r_A$ at **B** and ray $r_C$ at **C** a known time $\tau_S$ apart in the S frame. Counter **A** receives the rays a time $\tau_{LAB}$ apart. In practice, $y$ is infinitesimal, $y << \delta << x$, and $r_C$ and $r_A$ are parallel and of equal length.

The source S is given a known round-trip velocity $v$ but an unknown one-way velocity $v(\theta_B)$ between lab points B $(x, -y/2)$ and C $(x, y/2)$. The emitted radiation has an amplitude maximum at B and next at C, a known time $\tau_S$ later, as measured in the source rest frame, and a measured time $\tau_{LAB}$ later. $\tau_{LAB} = y/v(\theta_B)$, as measured by a counter/clock combination A $(0, 0)$ having an aperture $\delta$. In reality $y$ is infinitesimal, $y << \delta << x$, so the rays emitted at B and C are parallel to the X-axis.

With

$$\gamma = (1 - v^2/c^2)^{-1/2} \qquad (8)$$

SR predicts[18] $\tau_{LAB} = \gamma \tau_s$, therefore the frequencies are

$$\nu_{LAB} = \nu/\gamma. \qquad (9)$$

This is the transverse Doppler shift.

Winnie[7] has derived synchronization-dependent ('$\sigma$-dependent') Lorentz transformations which, independently of the value of $\sigma$, duplicate the standard SR kinematical conclusions about space-contraction, time-dilation, etc… Winnie proved that whatever a signal's one-way velocity may be, all SR predictions depend only on $v$, its round trip velocity: in Eq. (8), $\gamma$ must be computed using $v$ rather than the one-way velocity $v(\theta_B)$ and Drągowski_and_Włodarczyk's proposed experiment cannot measure one-way velocities. (Actually, Equation (7) gives the same result: in Figure 2, $r_C=r_A$, $\theta_C=\theta_A$, so $v(\theta_B)=v$, as one would expect from clause (*a*) of the SR Postulate.)

2.2.3 Dynamics and Electromagnetism.

Analyzing classical physics, Giannoni[8] has distinguished between $\sigma$-independent and $\sigma$-dependent quantities. The latter are Minkowski 4-vectors and 4-tensors with some components being $\sigma$-dependent and the others $\sigma$-independent. The displacement vector {*r*, *t*}, where *r* is $\sigma$-independent and *t* is $\sigma$-dependent, is a $\sigma$-covariant 4-vector and velocity, the time-derivative of the displacement, is $\sigma$-contravariant. Scalar products of $\sigma$-vectors and $\sigma$-tensors are $\sigma$-independent: they are measurable quantities, numerically equal to the standard SR predictions.

It has been suggested that one can measure a particle's $v(\theta)$ after it traverses a semi-circular path in a magnetic field by measuring its momentum, $p=m\gamma v$, before and after, $m$ = rest mass. Giannoni has shown that *mass* is as $\sigma$-dependent as $v(\theta)$, the $\sigma$-dependent velocity, while $\gamma$ is a scalar. In the particle's semi-circular motion, both mass and velocity change while $p$ remains constant: $p_{before}$ and $p_{after}$ yield only the round-trip velocity $v$.

Synchronization-dependence of mass is easily shown. Consider a $\pi^+$, $\pi^-$ pair produced by $K^0_S$ decay with total ***p***=0 in the pair center-of-mass system. The pions have velocities $v_\pm$ along the $x_\pm$ directions and, momentum $p_\pm=m_\pm\gamma_\pm v_\pm$, with $v_\pm=v/(1\mp\beta\sigma)$, Equation (4). Winnie[6] has shown $\gamma_\pm=(1-v^2/c^2)^{-1/2}$, is a $\sigma$-scalar. Since $p_+=p_-$ along the *x*-axis, we have

$$m_\pm=m(1\mp\beta\sigma). \qquad (10)$$

Mass is $\sigma$-covariant. For a free massive particle $E=mc^2$, becomes $E_\pm=m_\pm\gamma c^2$, and energy is $\sigma$-dependent. For a zero-mass particle $E^2=c^2p^2+m^2c^4$ reduces to $E_\pm = c_\pm p$, again $\sigma$-dependent.

Yet, consider a light-signal consisting of $N$ photons of frequency $v$. The total energy is

$$E=Nhv. \tag{11}$$

(as Einstein also suggested in 1905), Frequency is measured with only one clock. Thus, this expression is $\sigma$-independent. This glimpse into quantum mechanics suggests that $\sigma$-dependence is not intrinsic and that one-clock experiments that do not consider light-signals to be impenetrable black boxes may yield $\sigma$-independent values for some quantities, thus allowing measurement of $c(\theta)$ and synchronization. This is quite important for Quantum Mechanics.[12] The next Section considers the wave properties of signal-carrying light and presents a $\sigma$-independent method of measuring $c(\theta)$ and synchronizing clocks.

Any proposed experiment must show that the $\sigma$-independent quantities to be measured will yield a value for $\sigma$ or for a $\sigma$-dependent quantity, as in Equations (19), (20), (21), (23) and (24).

## 3. One-Way Measurement of Light's Wavelength and Velocity: Indications that the One-Way Velocity of Light is $c$.

For a wave $c= v\lambda$. With CS, $c$ becomes $c(\theta)$ and, for light of a frequency $v$,

$$c(\theta)= v\,\lambda(\theta). \tag{12}$$

This Section discusses the measurement of $\lambda(\theta)$. Isolated experiments show $\lambda(\theta)$ is isotropic. If after these experiments are re-analyzed or extended, they determine $\lambda(\theta)$ is isotropic along three non-coplanar axes, this would demonstrate isotropy in $c(\theta)$. These experiments use only one clock: synchronization is irrelevant.

*3.1 Determination of an Asymmetry in the Wavelength of Transmitted Light.*

Given Section 2.1, when light from a source at $x=0$ traverses two parallel slits at $x=s$ producing two light rays that recombine at $x=b$, the wavelength determined by the phase difference at $b$ between the two rays is independent of $c(\theta)$: this measurement does not yield an "average $\lambda$" in the +$x$-direction, but its round-trip wavelength. The measurement of wavelength proposed here does not rely not on the measurement of phase differences between rays but on the diminution, below a specific frequency, of the light intensity transmitted through a system. We seek a large diminution to ensure the transmission of light is a one-way process, rather than the superposition of multiple reflections within the system.

The definition of 'wavelength' as the distance, at a given time, between two successive peaks of the wave amplitude implies one has absolute synchronization. I propose instead an instrumentalist definition of wavelength: for a monochromatic beam impinging on a system that either detects light or acts as a filter one identifies a length $\Lambda$ in the system where one finds absorption or sizeable attenuation of the transmitted intensity and one determines how $\lambda(\theta)$ depends on $\Lambda$. One must verify that this absorption or attenuation depends on the wavelength rather than frequency by immersing the system in a fluid with a different index of refraction. Intensity can be measured by a single clock as the number of photons/sec and is $\sigma$-independent. The proposed method should be implemented simultaneously in opposite directions.

Since $\lambda = c/v$ for a round trip, $\lambda_\pm = c_\pm/v$ in a chosen direction and $\lambda_\pm$ can be obtained by measuring the wavelength-dependent diminution in the transmitted intensity when one interposes a variable-width transparent airgap in light's path between two glass blocks. I rely on Maxwell's electrodynamics which are subsumed under SR.

Replacing $c$ by $\lambda$ in Equations (2a - 2b) yields the direction-dependent $\lambda_+$ and $\lambda_-$ and

$$1/\lambda_+ - 1/\lambda_- = -2\sigma/\lambda. \tag{13}$$

While the CS thesis allows light of any frequency to have infinite velocity and therefore infinite wavelength in some direction, this is precluded by everyday observations: such light would never be attenuated when passing through an airgap between two glass blocks and, as already noted in 1982 [15] in this respect, such light would never pass through a pinhole. It also raises the question whether such light can visible to the eye oe be detected by similar size day detectors.

*3.2 The Wavelength of Light Transmitted Because of Frustrated Total Internal Reflection.*

Snell's law of refraction predicts that when visible light passes from a medium with a refractive index $n_1$ into one with index $n_2$, $n_1 \sin \phi_1 = n_2 \sin \phi_2$, with the angles measured with respect to an orthogonal to the interface. If $n_1 > n_2$ and $\phi_2 = \pi/2$ in the $n_2$ medium, there is a critical angle $\phi_c$, $\sin \phi_c = 1/n_1$, beyond which there is total internal reflection.

For a glass/air interface $\phi_c = 42°$ (for glass $n_1 = 1.50$, for air $n_2 = n_a = 1$). For $\phi_1 > \phi_c$, in order to satisfy the boundary conditions for the **E** and **B** fields at the interface, there is, beyond the interface, a surface (*"evanescent"*) wave propagating along the interface and, also, an attenuated transmitted wave through the interface[19, 20] with intensity ($E^2$) $I_T = KI_i e^{-\eta d}$ where:

$$\eta = (8\pi/\lambda_a)\{(n_1 \sin \phi_1)^2 - (n_2)^2\}^{1/2}, \tag{14}$$

$d$ is the distance light penetrates into the interface, $\lambda_a$ the round-trip wavelength in air, $K$ a constant, $I_i$ is the initial intensity, and $\eta = 4.44/\lambda_a$ at $\phi_1 = 45°$ for a glass/air interface. The intensity transmitted through a parallel airgap of width $\Lambda$ between two glass blocks is

$$I_T(\Lambda) = KI_I e^{-\eta \Lambda} = KI_I e^{-4.44\Lambda/\lambda a}, \tag{15}$$

an effect called *frustrated total internal reflection* ('FTIR') or *optical barrier penetration*. $I_T(\Lambda)$ decreases steeply for $\lambda a < 4.44\Lambda$. Equation (15) is valid for all $\Lambda$ for **E** normal to the plane of incidence but valid only if $\Lambda > \lambda_a$ for **E** parallel to that plane, where, for $\Lambda < \lambda_a$, $I_T$ is roughly constant. Henceforth, I consider only **E** normal to the plane of incidence.

*3.3 The CS Thesis and the Wavelength of Light Transmitted through an Airgap.*

Snell's law holds under the CS thesis, with $n_1$ and $n_2$ independent of $\sigma$.[15] Giannoni[8] has adapted Maxwell's electrodynamics to the CS thesis and shown that Maxwell's equations hold for light propagating in the $x_\pm$ directions. Equation (15) becomes:

$$\eta_\pm = (8\pi/\lambda_{a\pm})\{(n_1 \sin \phi_1)^2 - (n_2)^2\}^{1/2} \quad (16)$$

where $\eta_\pm = 4.44/\lambda_{a\pm}$, for $\phi_1 = 45°$ and a glass/air interface. With initial intensities $I_{1\pm}$ and **E** normal to the plane of incidence, the intensities transmitted through an airgap of width $\Lambda$ are:

$$I_{T\pm}(\Lambda) = KI_{1\pm} e^{-4.44\Lambda/\lambda_{a\pm}}. \quad \pm \quad (17)$$

With equal initial intensities $I_{1\pm}$, for a given $\Lambda$ the ratio $R(\Lambda)$ of the transmitted intensities is:

$$I_{T+}(\Lambda)/I_{T-}(\Lambda) = R(\Lambda) = e^{(-4.44\Lambda/\lambda_{a+})}/e^{-(-4.44\Lambda/\lambda_{a-})}. \quad (18)$$

Given Equation (13), $R(\Lambda) = e^{8.88\sigma\Lambda/\lambda_a}$ and $\sigma$ is:

$$\sigma = \lambda_a (\log_e R(\Lambda))/8.88\Lambda. \quad (19)$$

Measurement of the transmitted intensity through an airgap between two transparent blocks yields the one-way wavelength and velocity for light traversing the gap. The ratio of transmitted intensities in opposite directions yields $\sigma$, with $\sigma$ expressed in terms of $\sigma$-independent quantities.

Excellent agreement between Eq. (15) and experiment for $3\lambda_a < \Lambda < 8\lambda_a$ was found by Coon[21] (1966) using a monochromatic beam from a Hg lamp and a photomultiplier tube counting the number of photons transmitted through an airgap vs the gap's width for an airgap between two glass prisms: the observed $\lambda$ is always $c/v$. Thus, $c(\theta)$ for light traversing the Coon apparatus is $c$. Others[22, 23, 24] also have found that $c(\theta)$ for light traversing their apparatus is $c$. FTIR has also been observed with microwaves.[19]

*3.4 Proposed Experiment to Measure Simultaneously $\lambda_+$ and $\lambda_-$, and, Thus, $c_+$ and $c_-$.*

Figure 3 depicts a proposed apparatus, following Coon,[21] for a simultaneous measurement of $\lambda_+$, $\lambda_-$ and, thus, $c_+$ and $c_-$. The apparatus comprises two transparent blocks (index $n_1$) and a variable-width parallel gap filled with a transparent medium (index $n_{2<} n_1$) between the blocks.

One measures, one photon at-a-time, in opposite directions, the optical intensity transmitted through the system as a function of the gap-width $\Lambda$. $S_+$ and $S_-$ are sources of identical monochromatic constant-intensity parallel light beams with velocities $c_+$ and $c_-$, wavelengths $\lambda_+$ and $\lambda_-$, and **E** normal to the plane of incidence. At $\phi_1 = 45°$ with respect to the gap, are the incident, transmitted, and reflected intensities, $I_{I\pm}$, $I_{T\pm}$, and $I_{R\pm}$, respectively. Photomultiplier tubes $P_{T\pm}$ and $P_{R\pm}$ detect the transmitted and reflected photons, and $P_{e\pm}$ monitor the evanescent waves.

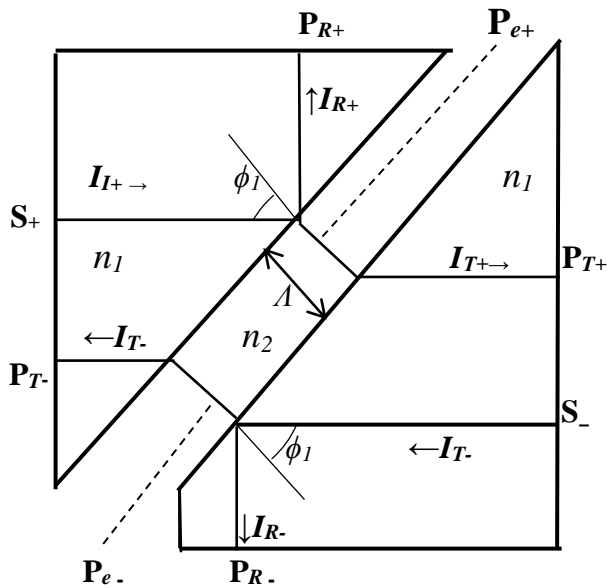

Figure 3. <u>Bi-directional measurement of the directional dependence of the transmission of light through a gap.</u> Two transparent blocks (index $n_1$) have a variable transparent gap $\Lambda$ (index $n_2$) between them. $\mathbf{S_\pm}$ are identical light sources and $\mathbf{P_{T\pm}}$, $\mathbf{P_{R\pm}}$, and $\mathbf{P_{e\pm}}$ are photomultiplier tubes measuring the transmitted, reflected, and evanescent intensities.

Experimenters should use media with different $n_1$ and/or gaps with different $n_2$ and verify that the same value of $\sigma$ is obtained. A determination that $c(\theta)$ is isotropic requires simultaneous bi-directional measurements along three non-coplanar axes. The small bulk of the apparatus makes this method especially appropriate. This experiment can address other theories regarding the velocity of light, Anderson et al.[14]

## 4. On Thyssen's Claim that Distant Simultaneity is Conventional: Synchronization and the One-Way Velocity of Light from Relativistic Time-Dilation.

Once the value of $\sigma$ is established, Equation (19), one can synchronize distant clocks using Equation (4), $c(\theta)= c/(1-\sigma \cos\theta)$. I shall provide below further indications that $c(\theta)=c$.

This negates Thyssen's claim[2] that "the relation of distant simultaneity is conventional because it is unverifiable. Even if the relation of distant simultaneity really exists, we nevertheless fail to have epistemic access to it, and are thus forced to treat this notion in a conventional manner." I present below a synchronization method derived directly from SR.

*4.1 Synchronization from Relativistic Time-Dilation with a Moving Clock.*

As discussed in sub-Section 2.2.2, SR predicts that if a clock in a rest frame measures a time $t_S$ (its 'proper time') for travel with velocity $v$ between lab-points A and B, the travel time $t_{LAB}$ that would be measured by synchronized clocks at A and B, with $\gamma=(1-v^2/c^2)^{-1/2}$, is $t_{LAB}=\gamma t_S$ and Winnie[8] proved $t_{LAB}$ is independent of $\sigma$. Lacking synchronized clocks, we proceed as follows.

The method presented here is a *fast clock transport* method, the obverse of the 'slow clock transport' method introduced by Ellis and Bowman[25] and criticized by Grünbaum[26] and others.

Figure 4 shows, on the X-axis of an inertial frame, a clock/computer combination S at $x<0$ together with identical clock/computers A at $(0,0)$ and B at $(X,0)$. Clocks in S, A, and B are light-clocks[27] where a light pulse travels back-and-forth between mirrors a distance $L$ apart. The clocks' fundamental frequency is $v_L=c/2L$, and each clock emits a signal at that frequency.

(1) At rest, the clocks' isochrony is verified by exchanging signals of frequency $v_L$.

(2) Next, S assumes a uniform round-trip velocity $v$ towards A and B such that $L$ is unchanged, leaving S isochronous with A and B. Preservation of isochrony between clocks S, A, and B when S is traveling is derived directly from the R-modified SR Postulate --- the round-trip speed of light is the same for all inertial observers.

(3) When S passes A, S initializes its clock and signals A to initialize its clock, $t_S=t_A=0$.

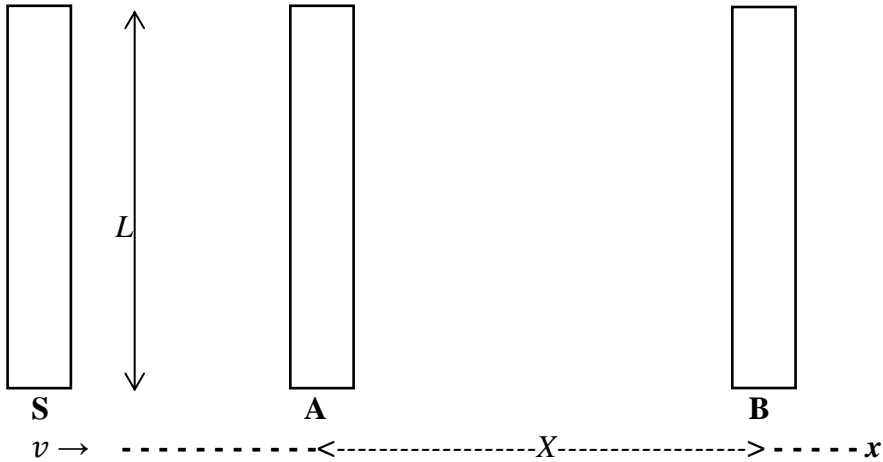

**Figure 4**. <u>Synchronization with a Moving Clock</u>. A and B are isochronous counter-clock combinations. In sub-Section 4.1, S is a moving counter-clock isochronous with A and B. In sub-Section 4.3, S is a burst of unstable particles.

(4) As S travels towards B with round-trip velocity $v$, A, B, and C measure $v$ from the Doppler-shifted frequency $v_S$ of the $v_L$ signals between S and A and S and B and all compute $\gamma$,

(5) As S passes B, S signals to B the reading of clock S, $t_S$.

(6) Given Equation (8), B sets its clock to read

$$t_B = \gamma t_S. \qquad (20)$$

(If there is a delay $\Delta$ when the S signals reach A and B in steps (3) and (5), this delay should not affect the time difference between clocks A and B, especially if $\Delta < c/2L$ and S grazes them. Otherwise, this raises the question whether one can synchronize clocks at rest abutting each other. In sub-Section 4.3, below, the moving clock goes through the fixed clocks!)

All the quantities used in deriving Equation (20) are $\sigma$-independent and <u>clocks A and B are now synchronized in a conventionality-independent manner</u>, resulting in Einstein synchronization. This method can be used with any clock S the isochrony of which is unchanged after it has been accelerated.

If S reverses direction after (5) it verifies the isotropy of this synchronization by verifying that the clock readings difference $t_B - t_S = \gamma t_S$ on the return trip.

The above method allows the synchronization of an arbitrary number of clocks in all directions.

Given the *isochrony* of clocks S, A, and B, the proposed method, without any assumptions about one-way velocities, allows conventionality-free *synchronization* of several clocks at arbitrary locations on an inertial frame. Both Einstein and Reichenbach could have adopted it.

*4.2 Independently of Synchronization, This Method Yields the One-Way Velocity of Light.*

After step (4) S can measure $c(\theta_x)$, $c(\theta)$ along the $+x$-direction with only a marker M is placed at a distance $X$ from A. As S passes M it measures $t_S$, and sets

$$c(\theta_x) = X/\gamma t_S. \tag{21}$$

Similarly, one can determine $c(\theta_y)$, $c(\theta_z)$, and, also, $c(\theta)$ along the negative directions. If all six of these $c(\theta)$ are equal to $c$, <u>the one-way velocity of light is $c$</u>.

*4.3 Present-day Achievable Synchronization from Relativistic Unstable-Particle Bursts.*

Consider again Figure 3 where S is now a well-collimated unstable-particle burst exiting an accelerator with particles of lifetime at rest $\tau$ and velocity $v_{SEL}$ selected by traversing crossed electric and magnetic fields. <u>$v_{SEL}$ is a round-trip velocity,</u> Giannoni.[8] The burst then traverses isochronous counter/clock combinations A (0, 0) and B (X, 0) that determine the number of traversing particles $N_A$ and $N_B$ at A and B respectively. $N_A$ and $N_B$ are the same in the burst and lab frames.

The number of particles in the burst $N_S(t_S)$ decreases as $e^{-t_S/\tau}$, $t_s$ measured in the burst rest frame. $N_S(t_S)$ constitutes a moving clock. As S passes A, A sets its clock at $t_A = 0$ and A broadcasts to B the number of particles $N_A$ passing A. As S passes B, B determines the number of particles $N_B$ traversing B, which equals $N_A\, e^{-t_S/\tau}$, then computes $\Pi(X) = N_A/N_B$ and

$$t_S = \tau\, \log_e \Pi(X). \tag{22}$$

Given time-dilation, the travel time from A to B in the lab frame is $t_{LAB} = \gamma t_S$ or:

$$t_{LAB} = \gamma \tau\, \log_e \Pi(X), \tag{23}$$

where $\gamma$ is computed from $v_{SEL}$. B sets its clock to read $t_{LAB}$, and the <u>two clocks are now synchronized in a synchronization identical to Einstein's</u>.

Similarly, one can synchronize several clocks $B_i(x)$ on the x-axis, and on other beam-lines so that an arbitrary number of clocks are synchronized along all spatial directions.

*4.4 The One-Way Velocity of Light Without Clock Synchronization.*

For all angles above, $t_{LAB} = \gamma\, \tau\, \log_e \Pi(X)$ and the <u>one-way velocity</u> $v_{LAB}(\theta) = X / t_{LAB}$, is:

$$v_{LAB}(\theta) = X\, (\gamma\, \tau\, \log_e \Pi(X))^{-1}, \tag{24}$$

$v_{LAB}(\theta)$ is expressed in terms of σ-independent quantities and determined without the use of clock synchronization.

Measurements of $\tau$, such as Rossi and Hall [28] and Durbin et al.[29] are often stated in terms of the *mean decay-length* $\chi = \gamma\, \tau\, v_{LAB}(\theta)$ or

$$\chi = X(\log_e \Pi(X))^{-1}. \tag{25}$$

A pulse of $\pi^+$ particles, $m = 140$ MeV/$c^2$, $\tau = 140$ ns (easily measured with a single clock),

energy ≈ 1 GeV, $\gamma \approx 7$, $\beta=0.99$, $\chi \approx 300$ m, is a suitable burst S for implementing the proposed synchronization method. Measurements along three non-coplanar beamlines are necessary to determine whether $\sigma$ is isotropic.

*4.5 Where the CS Thesis Stands.*

Subsections 4.3 and 4.4 allow a thorough examination of the validity of the CS thesis which predicts $\sigma$ can be chosen at random, $0 \leq \sigma \leq 1$ in Equations (4-5).

Specifically, the above treatment through Equation (24) applies under CS. $N_S(t_S)$ is a $\sigma$-independent clock. $X$, $\gamma$, and $\tau$ are all σ-independent and so must $t_{LAB}$ be. Therefore:

(1) clocks can be synchronized in a convention-free manner, contradicting conventionality;

(2) one-way velocities can be measured;

(3) one-way velocities and synchronization are independent;

(4) $v_{LAB}(\theta)$ is the same for each burst S traveling with the same $\gamma$ along any beamline.

This last item is supported by ample empirical evidence. Experimenters plan experiments according to the expected mean decay-length $\chi = \gamma\tau v_{LAB}$ using $v_{LAB}(\theta) = v_{SEL}$, the round-trip velocity. If $\gamma=10$ ($\beta=0.995$), and $\sigma=0.2$, CS predicts only a 20 percent probability that experiments will show $0.83 v_{SEL} < v_{LAB}(\theta) < 1.25 v_{SEL}$. Experimenters have not reported observing $v_{LAB}(\theta) \neq v_{SEL}$.

Equation (5) predicts a $\beta\sigma\cos\theta$ dependence for $v(\theta)$. This can be explored by comparing the values of $v_{LAB}(\theta)$ obtained simultaneously along different beamlines at a given accelerator or on a given beamline over several weeks during which time the beamlines partake of the earth's rotation and heliocentric revolution. (The same is true for the experiments discussed in subsection 3.3.). Such research may address questions about the velocity of light raised by Mansouri and Sexl.[30]

## 5. Conclusion.

The one-way velocity of light is measurable. Using optical barrier penetration, one can measure the one-way wavelength of light and, hence, its one-way velocity. Also, the mean decay times of unstable particles allow measurement of the particles' one-way velocity. This measurability itself invalidates Reichenbach's thesis. For both cases, all available data indicate isotropy in these one-way velocities, thus supporting standard SR, but one must ensure isotropy along three non-coplanar axes. Also, with a moving clock isochronized with two or more stationary clocks, one can synchronize the stationary clocks without knowing any one-way velocity: with SR, isochrony yields synchronization. Bursts of unstable particles are appropriate moving clocks. Therefore, clock synchronization and one-way velocities are independent matters and, concerning each, Reichenbach's Conventionality of Synchronization thesis is invalid.